\documentclass[a4paper,12pt]{article}
\usepackage[utf8]{inputenc}
\usepackage[english]{babel}
\usepackage{amsmath}
\usepackage{amsfonts}
\usepackage{amssymb}
\usepackage{enumitem}
\usepackage{float,natbib}
\usepackage{graphicx}
\usepackage{geometry}
\usepackage{caption}
\usepackage{subcaption}
\usepackage{url,color}
\usepackage{booktabs}
\usepackage{dcolumn}
\usepackage{hyperref}
\usepackage{lscape}

\graphicspath{ {images/} }
  \usepackage[titletoc,toc,page]{appendix}
  
\title{\Large Measuring Social  Well Being in The Big Data Era: \\ Asking or Listening?}
\author{\normalsize Matteo Curti
		\thanks{Academic Affairs Division, Bocconi University}
		\and
		\normalsize Stefano Maria Iacus 
		\thanks{Department of Economics, Management and Quantitative Methods, Universit\`a degli Studi di Milano}
            \and
       	\normalsize Giuseppe Porro
       	\thanks{Department of Law, Economics and Culture, Universit\`a degli Studi dell'Insubria}
            \and
        \normalsize Elena Siletti
        \footnotemark[2]
}

\date{}
\begin{document}
\maketitle

\paragraph{Abstract:}

The literature on well being measurement seems to suggest that ``asking'' for a self-evaluation is the only way to estimate a complete and reliable measure of well being. At the same time ``not asking'' is the only way to avoid biased evaluations due to self-reporting. Here we propose a method for estimating the welfare perception of a community simply ``listening'' to the  conversations on Social Network Sites. 
The Social Well Being Index (SWBI) and its components are proposed through to an innovative technique of supervised sentiment analysis called iSA which scales to any language and big data.
As main methodological advantages, this approach can estimate several aspects of social well being directly from self-declared perceptions, instead of approximating it through objective (but partial) quantitative variables like GDP; moreover self-perceptions of welfare are spontaneous and not obtained as answers to explicit questions that are proved to bias the result. As an application we evaluate the SWBI in Italy through the period 2012-2015  through the analysis of more than 143 millions of tweets.

\paragraph{J.E.L. Classification:}
I30; I31

\tableofcontents
 
\section{Introduction}
\label{intro}
The measurement of well being is a matter that psychologists, social scientists and policymakers have been tackling for decades. In fact, personal well being is commonly considered a topic for psychological science, because it concerns the subjective feeling of individuals: on the contrary, social well being is actually a collective dimension of a community and its value is considered a synthetic and significant description of how the development of a socio-economic system is well balanced and sustainable.

Here we review the methodologies that have been applied to measure well being as a social dimension: their purposes and limitations of the different approaches will be analyzed and their progressive evolution described. Consequently, we make a new proposal for measuring social well being, relying on the availability of big data sources provided by the web-based social networks, and on one of the most recent techniques for sentiment analysis. This approach disentangles the main methodological issues raised in the literature on well being measurement, and produces a set of indicators that span the wide range of well being perceptions.

The paper is structured as follows: the traditional measurement of well being through GDP is discussed in Section \ref{sec.1}, while the concerns raised by the so-called capability approach are presented in Section \ref{sec.2}. The following Section \ref{sec.3} examines the contribution by the Stiglitz Comnmission, that formalizes the multidimensional measure of well being: the best known multivariate indicators of social well being are described. Section \ref{sec.4} analyzes the evolution towards the so-called survey approach and describes the main surveys available and the studies they have generated. Section \ref{sec.5.1} introduces the novelty of social networks and the repositories of information they make available: the relationship between these big data sources and well being measures is discussed and previous researches on the subject are presented. The new set of well being indicators and their synthetic representation through a Social Well Being Index (SWBI) are described in Section \ref{sec.5} and its application to the Italian case for the period 2012-2015 is presented.  Section \ref{sec10} contains the detailed description of iSA, the sentiment analysis technique applied to construct the indicators though social networks sites. Finally, Section \ref{concl} concludes the paper.

\section{GDP: Development, Usage and Limitations}
\label{sec.1}
GDP had a widespread success as an index measuring socio-economic progress, in spite of its well-known shortcomings as an indicator of well being: in 1934 Kutznets, Nobel Prize Laureate in 1971 and creator of the modern concept of GDP said - in that moment he was chief architect of the United States national accounting system - at the US Congress that ``the welfare of a nation can scarcely be inferred from a measurement of national income'' \citep{kuznets_1934}. The reasons why GDP had such a success are its capacity to connect goods and services with different nature thanks to monetary valuation \citep{stiglitz_etal_2009}, its linear methodology, objectivity and clearness (for example in public debates) and the usefulness in international comparisons. 
GDP is, by definition, an aggregate measure of production and it includes the production of all final or collective goods or services that are supplied to units other than their producers, in a certain period of time as summarized by the  \citet{insee_2008}. 
Although its correlation with many indicators of living standards is high and significant, this correlation is not universal \citep{stiglitz_etal_2009} and differences in income explain only a low proportion of the differences in happiness among persons \citep{frey_stutzer_2002}: indeed GDP is criticized for being a lacking indicator of social welfare and therefore a misleading guide for public policies \citep{fleurbaey_2009}. 
Economists, psychologists and philosophers have become increasingly interested in self-reported measures of well being, their significance and their usefulness in policy-making \citep{deaton_stone_2013}. \cite{sen_2003} warned that a focus on economic growth, and on GDP, as a measure of human development is a mistake, since there is not a one-to-one correlation with growth and quality of life and too much emphasis on GDP as a unique benchmark of well being can lead to wrong considerations and policy decisions: it is a measure of market production, more useful to measure the aggregate supply side of economies than living standards of citizens \citep{stiglitz_etal_2009}. For example, since the second world war GDP in many countries has risen, but self-reported subjective well being has not increased or has fallen \citep{frey_stutzer_2002}. In 1974 Easterlin raised a paradox that became a puzzle for economists and social scientists: beyond a certain level of income (i.e. GDP per capita), enough to meet basic needs, the relationship between income and happiness are little, if any \citep{choudhary_etal_2012} anf, further, there is a virtually no gain in human development and well being from annual income beyond \$\,75,000 \citep{kahneman_deaton_2010}. For sure, income is generally an important means to well being and freedom, but it can only be used as a mere proxy for what really matters for people\footnote{March 18, 1968, Robert F. Kennedy at the University of Kansas said: ``Yet the gross national product does not allow for the health of our children, the quality of their education or the joy of their play.  It does not include the beauty of our poetry or the strength of our marriages, the intelligence of our public debate or the integrity of our public officials.  It measures neither our wit nor our courage, neither our wisdom nor our learning, neither our compassion nor our devotion to our country, it measures everything in short, except that which makes life worthwhile''.} \citep{robeyns_2005}.
Many issues have been raised about the adequacy of GDP as a well being indicator. First of all, monetary evaluation of market transaction is the starting point in measuring economic performance and so prices are fundamental, but they may not exist for some goods or services or they may not represent the real value for the society. Then, only goods and services that are officially exchanged on markets are included in GDP calculations, while others, relevant in achieving higher level of well being, are overlooked. It is the case of common goods provided by governments like security, freedom and democracy, but also volunteering activities and social relations, health and longevity and all the spectrum of shadow economics such as house work and services provided by family members.
A main remark, commented by \cite{stiglitz_etal_2009}, concerns the change in quality of goods and services: it is a dimension vital to notice, but extremely difficult to measure (generally complex and multi-dimensional). Secondly, market power and the effect of imperfect competition: firms with strong market power can rise prices, and their profits, generating a loss of consumer surplus that households face \citep{stiglitz_etal_2009}. GDP notices the unilateral rising in prices, while the consumer drop in wealth is overlooked. Third, aggregate production measures do not take into account negative externalities such as the degradation of natural environment and resources and ``assets'' (like environment and biodiversity) and pollution.
Finally, using GDP as a proxy of well being can lead to misleading observations concerning wealth of societies. On the one hand negative events such as natural disasters, earthquake or floods, or big car accidents, reduce wealth of society but can increase GDP, on the other hand it doesn't take into account distribution of income, so that a great disparity and poverty of capabilities wouldn't be noticed. Whether economists and social scientists have tried to correct GDP adding or subtracting monetised aggregates reflecting different kind of quantities and dimensions \citep{nordhaus_tobin_1973}, alternative approaches to the study and measurement of well being and better indicators of social welfare have been developed across the years: they are a core issue in the public debate and a primary concern for policy-makers. One of the most influential approaches to the analysis of the human wealth is the capability approach, developed, between the 80s and the 90s by the economist and philosopher Amartya Sen.

\section{Capability Approach}
\label{sec.2}
\subsection{Definition, Concepts and Impact}
\label{sec.2.1}
The capability\index{capability} approach was developed by the Indian Nobel Amartya Sen\index{Sen}. The capability approach, which has its roots in Aristotle, Smith and Marx, is a broad normative framework for evaluation and assessment of individual well being and social arrangements, the design of policies and proposal about social change \citep{robeyns_2000, robeyns_2006}. It provides a framework to carry out exercises in welfare comparisons. \citet{sen_2008} has stressed ``the plurality of purposes for which the capability approach can have relevance'' but, although it is used in a wide range of fields, such as development studies, welfare economics, social policy, political philosophy etc., it doesn't tend to find an explanation of social phenomena, an aim that requires additional social theories, but it helps in their conceptualization and evaluation \citep{robeyns_2005, robeyns_2006}. The capability approach covers all dimensions of human well being: it regards development, well being and justice and pays much attention to the linkages between material, mental and social well being. 
One of the main feature of the approach is the distinction between means and ends of human action. Such a distinction introduces two central concepts: functioning and capabilities, realized and effectively possible. The former represents the achievement of a person, it points out the thing a person manages to do or to be in leading a life, his or her ``doings and beings''. Functioning are, more directly related to living conditions, and more empirically measurable, since they concern different aspects of living conditions \citep{sen_hawthorn_1988}. Differently, capabilities are possibilities, chances, the alternative combinations of functioning that one can choose to achieve and pursue in his or her life. Capability is thus a set of vectors of functioning, reflecting the person's freedom to lead one type of life or another \citep{sen_1992,sen_2008}, and, as a sort of degree, barely detectable and quantifiable.
The capability approach has also had a relevant and long-lasting political impact: since 1990 UNDP adopted it in elaborating its Human Development Index (HDI) and the annual Human Development Report \citep{robeyns_2006}. The HD Index has been developed by Amartya Sen and the economist Mahbub ul Haq. Its theoretical background lie on the theory of human development by the Indian philosopher. HDI, in measuring  eing, takes into account three dimensions: health, education and material standards of living. The proxies used (for a total of four) are: for first domain life expectancy at birth, for the second the mean of schooling years and the expected years of schooling, while the third is estimated through Gross National Income per capita (PPP \$). In its 2014 edition the calculation is made in two main steps. At first, dimensional indices are created and normalized in a (0-1) interval \citep{hdro_2014}. Then, HDI is built as the geometric mean of the three dimensional indices. In this way, countries are classified in quartiles, according to their score: very high, high, medium and low human development.

\subsection{Capabilities for Measuring Well Being}
\label{sec.2.2}
Functioning and capabilities represents the informational base of this approach. It differs very much from other theoretical frameworks aimed at evaluating well being which use different informational focuses, for example on personal utility (pleasure\index{pleasure}, happiness\index{happiness}, desire\index{desire} fulfilment), opulence (income\index{income}, commodity bundles), means of freedoms\index{freedom} (holding primary goods or resources). But which are the links between well being and capabilities? As \citet{veenhoven_2010} points out, both at individual and social level emerges an interdependency between the two dimensions, even if relations and causality are less abundant and more difficult to assess from the capabilities' point of view \footnote{In its works, she identifies happiness with a four-fold taxonomy which stretches from hedonics, instantaneous feeling of joy, to person's life satisfaction, with a part of it or taken as a whole. In assessing the link between capabilities and happiness she defines happiness as ``overall life satisfaction'' and uses self-reported measure of happiness from World Database of Happiness.}.
Capability approach is based on a view of living as a mix of beings and doings, where quality of life could be measured in terms of capability to achieve valuable functioning. In doing this, it goes beyond ``commodity fetishism'', commodity-centrism and a mere attention on materiality. Nevertheless, material conditions and other objectively measurable dimensions of living have to be taken into account in evaluating well being: focusing only on subjective perspective could be very misleading, since it may fail to depict a person's real deprivation of material goods. 
As its name suggests, the capability approach sees capabilities as the primary informational base. Focusing on capability in evaluating well being doesn't seem to imply any informational loss. It allows considering not merely well being achievements, but also well being freedom (even if in detriment of its empirical implantation, as we will see later). Contrary to the standard consumer theory, where a set of feasible choices is assessed in relation of the best one available, in the capability approach the freedom to enjoy various possible beings and doings may have value in itself and an intrinsic importance for personal well being \citep{sen_2003, sen_2008}. In doing this, the capability approach does not impose a particular notion of good life and it is easy to identify its influence on the concept of luck-egalitarianism: each person should have the same real opportunity (capability), but individuals should be held responsible for their own choices \citep{robeyns_2006}.
But what capabilities count for well being? the real goal is to identify functioning and capabilities that value or which do not, what helps in pursuing well being and in human development and what does not: according to \cite{sen_2003} it is a reflective exercise, open to doubts and dialogue, but, in defining components of well being, ``it is undoubtedly more important to be vaguely right than to be precisely wrong''. For example, valuable functioning can be very different, in relations to the conditions where the person live, i.e. in developing or developed areas: some of them concern basic needs, such as being adequately nourished, being in good health, and other are more complex, like achieving self-respect and being socially integrated.

\subsection{Nussbaum Version of the Capability Approach}
\label{sec.2.3} 
Another version of the capability approach has been developed by Martha Nussbaum. According to her, freedom of choice requires for sure a formal defence of basic liberties, (i.e. rights on paper) and the assurance of certain levels of material conditions and real circumstances (the capabilities in Sen's approach). Nevertheless, for a concrete effectiveness of empowerment and human development another aspect has to be pointed out: the inner state of readiness to act, those set of ``central human capabilities'' that allows people to transform chances into outcomes, means into ends \citep{nussbaum_2000}. At this point it is clear that Sen and Nussbaum use a different notion of capabilities. In the former's works this concept refers to a real or effective opportunity, while in the latter's ones it pays more attention to people's skills and personality traits \citep{robeyns_2005}. Nussbaum's interpretation of the term capabilities as human attitudes recalls the concept of ``degree of conversion'' introduced by \cite{sen_2003} to indicate the different degree with whom people can transform commodities in capabilities and functioning, particularly those which \cite{robeyns_2000, robeyns_2005} identifies as ``personal'' and ``social''. Nussbaum elaborates a list of three types of conversion factors:
\begin{itemize}
\item Personal: features of a person that influence the conversion of the commodity into a functioning;
\item Social: values and institutions of the society in which one lives as well as social and public policies;
\item Environmental: concerning climate, natural characteristics, and geographical location.
\end{itemize}
In this perspective, Nussbaum elaborated a list of cross-cultural capabilities which refers to inner predispositions and personal skills, like (in \cite{veenhoven_2010} a synthesis): school intelligence, social intelligence and social functioning, leisure skills and moral competence\footnote{In their original version in \cite{nussbaum_2000}: life, bodily health, bodily integrity, senses - imagination - thought, emotions, practical reasons, affiliation, relations other living species, play, control over one's environment.}.

\subsection{Limitations of the Capability Approach}
\label{sec.2.4}
The capability approach have been implemented in a variety of contexts and fields as general framework, usually supported by other theories. \cite{robeyns_2006} makes a list of nine types of capability applications: general assessment of the human development of a country, poverty and well being assessment in advanced economies, theoretical and empirical analyses of policies, the assessment of small scale development projects, identification of the poor in developing countries, critiques on social norms, analyses of deprivation of disabled people, the assessment of gender inequalities and the use in descriptive or exploratory research (the first four of which appears here more relevant).
Even though the capability approach has been subject to great attention from researchers, policy makers and public actors and even if it represents a realistic alternative to traditional methods (like measurement of income or cost benefit analysis), its applicability continues to be an issue in the debate \citep{robeyns_2006} and doubts concerning the possibility to make empirical use of this ``richer but more complex procedure''  persist \citep{sen_2008}.
A great theoretical question concerning capability approach is the selection of capabilities: which capabilities and functioning are valuable in order to assess well being? The selection of valuable capabilities is a not-easy-to-pass step in the application of capability approach. Sen and Nussbaum adopted a different strategies in doing it. The first has never explained how such a selection could be done. Actually the main capabilities are chosen by researchers or social scientists in order to lead researches or produce official statistics: according to Sen the selection of capabilities is not the task of the theorist \citep{robeyns_2005}, but in its works he is not clear about the most suitable way. Even if it should be an ``act of reasoning'', taking into account the ends of the measurement  and the features of the reality that is assessed (developing or developed economy), it could be carried out by a democratic process \citep{robeyns_2006}. The Nobel Laureate \footnote{Amartya Sen was awarded the Nobel Memorial Prize in Economic Sciences in 1998} does not say anything more on this process, and according to some theorists, such as Robeyns, this vagueness is coherent with the framework-of-thought nature of the capability approach. On the other hand, Nussbaum tries to elaborate a cross-cultural normative account of capabilities related to the very human nature and to embed them in a philosophical framework recalling Aristotle, Marx and political liberalism\footnote{The word ``liberal'' in political philosophy refers to a thinking tradition that values the freedom of the individual and it shouldn't be confused with the same term used in everyday political life: it doesn't refers necessarily with left or right, nor it doesn't imply social or economic policies.}, in order to reach as wide consensus as possible on them among people with different concepts of a good life \citep{nussbaum_2000}. In spite of those efforts, capability approach has been often criticized for being not very democratic and suspected of paternalism.
Another issue has been raised by \cite{veenhoven_2010}, one of the main supporter of the happiness\index{happiness} approach in well being research, who states that the concept of capability, in spite of its focus on measurable and quantitative domains, is too wide and vague, leading to lose sight of the interrelation between environmental dimensions and individual skills, due to a lack of normative and theoretical background in studies and applications \citep{robeyns_2006}.
The practical usability of the capability approach faces some operative and procedural problems. The indefinite borders of the concept of capability often leads to a hard distinction between functioning and capabilities \citep{fleurbaey_2009}. Even if, from a theoretical perspective, this difference is clear in all works referring to the capability approach, at a more practical level it becomes less evident. The most part of implementations are based on functioning because they are more quantitatively measurable and allow for using available data from existing dataset, for example the ones collected by governmental agency and institutions, without distinction whether these data refer to capabilities or functioning \citep{fleurbaey_2009, robeyns_2006}. But the point is not only a methodological or observability concern, it refers also to the relationship between well being and achievements or opportunities. If what counts are capabilities, i.e. because they stress the importance of freedom and choice, is a composite index, built on functioning-data, a real measure of well being and coherent with its theoretical underpinnings?
At the end, even if one may be not so sure about the complete usefulness of the capability approach, as \cite{ellman_1994} seems to be, the question raised by \cite{sugden_1993}: it is plausible? is the capability approach a realistic way to measure well being? This question will probably remain without an answer. Nevertheless, academic and institutional actors are more and more engaged in developing alternative ways to deal with well being measurement. An example of such an effort is the work of the Commission on the Measurement of Economic Performance and Social Progress initiated by the French Government and chaired by Joseph Stiglitz.

\section{Stiglitz Commission Report: Measuring Well Being Through Different Dimensions}
\label{sec.3} 
In 2009, the so-called Stiglitz Commission (which also includes Fitoussi and Sen) observed that GDP shouldn't be dismissed and proposed to build a complementary statistical system, centred on people's well being and suitable for measuring sustainability, composed by a wide set of indicators\footnote{Including GDP, ``because no single measure can summarize something as complex as the well being'' (\cite{stiglitz_etal_2009}). Moreover, they sustained that synthetic indices could lead to a loss of information, as well as arbitrary assumptions in the weighting that has to be applied to the different dimensions and their sub-elements to arrive at a single index figure.}, quantitatively measured and representing both objective and subjective assessment of well being: including also people's perception of their quality of life. They identify key aggregated dimensions that should be taken into account: material living standards, health, education, personal activities including work, political voice and governance, social connections, environment and insecurity. With its work, the Commission made a sort of ``paradigmatic'' choice that have had a strong influence in further well being literature and, above all, practice.
Indeed, recently, following this path, statisticians and social scientists, universities and think tanks, governments and international organizations \citep{fleurbaey_2009}, have developed a huge number of well being indicators, with different structures (both composite indicator and dashboard of indices), considering a great variety of dimensions and for many purposes. Shown below some of these indices.

\subsection{Better Life Index - BLI}
\label{sec.3.1}
BLI and the ``How's Life'' report are the output of the Better Life Initiative launched by the OECD in 2011. BLI proposes a dashboard of 11 dimensions ``that can be considered universal'', i.e., Civic Engagement, Community, Education, Environment, Health, Housing, Income, Jobs, Life Satisfaction, Safety, and Work-Life Balance and 24 proxies. Those dimensions have the same weight: the BL Index has been conceived as an interactive tool that allows users, thanks to its web platform, to mix the set of proposed indices, giving different weights, in order to elaborate a index coherently with one's preferences \citep{oecd_2013}.

\subsection{Happy Planet Index - HPI}
\label{sec.3.2}
Developed by the \citep{nef_2012}, a British think tank, for the first time in 2006, and today at its third (2012) version, the HP Index uses data on life expectations, experienced well being (from the Gallup World Poll) and ecological footprint to produce an original overview on well being. The HP Index is calculated multiplying experienced well being with life expectancy in order to obtain a middle indicator of ``happy life'' which is then divided by ecological footprint.

\subsection{Fair Sustainable Well Being (Benessere Equo Sostenibile) - BES}
\label{sec.3.3}
BES\footnote{\url{http://misuredelbenessere.it}} or  ``Benessere Equo Sostenibile'' (Fair Sustainable Well Being) is the well being index elaborated by the Italian Institute of Statistics (ISTAT) setting up from a dashboard of twelve dimensions: Economic well being, Education, Environment, Health, Landscape and Culture, Politics and Institutions, Research and Innovation, Safety, Service Quality, Social Relations, Subjective well being, Work and Work-Life Balance. Although its clear conceptual and statistical similarity with the BLI, BES does differ from this and other examples shown above in denying any form of aggregation: in the periodical report, for each dimension, the entire set of proxies is presented and discussed. The ISTAT does not provide any form of aggregation inside and between dimensions composing BES, although some regional agencies like IRES Piemonte (the Regional Institute for Economic and Social Research of Piemonte) do.
IRES\footnote{\url{http://www.regiotrend.piemonte.it/qualita-vita/cruscotto-italia.html}} elaborates general and domain-specific composite indicators for Italian regions, presenting them with a format similar to BLI by OECD: avoiding any form of weighting but giving to the citizen the possibility of creating the index that reflects his or her own preferences.  

\subsection{Canadian Index of Well Being - CWI}
\label{sec.3.4}
Launched in 2011 with the first report on Canadian well being, it has been developed by the University of Waterloo. Similarly to some previous cases, CIW is a composite single-number indicator calculated as the arithmetic mean of eight domain values: Community vitality, Democratic engagement, Education, Environment, Healthy populations, Leisure and culture, Living standards, Time use and composed by eight normalized indicators each \citep{michalos_etal_2011}.

\subsection{Gross National Happiness - GNH}
\label{sec.3.5}
The concept of Gross National Happiness (GNH) was introduced in the 60s by the former King of Buthan Jigme Dorji Wangchuck\footnote{He was the Third Druk Gyalpo of Bhutan. He began to open Bhutan to the outside world, began modernization, and took the first steps toward democratization.} and then articulated, more recently, by his successor Jigme Singye Wangchuck\footnote{He was the King of Bhutan (Druk Gyalpo) from 1972 until his abdication in favour of his eldest son, Jigme Khesar Namgyel Wangchuck, in 2006. He is credited with many modern reforms in the country.}. Gross National Happiness has been defined as the degree to which citizens in a country enjoy the life they live and as every citizen counts equally much in this sum, the concept can be quantified using the average of individual happiness in the country \citep{veenhoven_2004}. It is then an aggregate index developed from thirty-three indicators, categorized under nine dimensions equally weighted: Community vitality, Culture, Ecological diversity and resilience, Education, Good governance, Health, Living standards, Psychological wellbeing, Time use \citep{bhutan_2015}. For every indicator it is identified a sufficiency cut-off, usually set higher than the poverty line (for some indicators it is set at the top level). GNH index is calculated using the following formula:
$GNH=1-(H\cdot A)$
Where $H$ represents the percentage of people who do not enjoy sufficiency cut-off in six or more domains, while $A$, the average proportion of domains in which people who are not happy (those who have not achieved the sufficiency cut-off in 6 out of 9 domains) have a shortage. An interesting feature of GNH is that any individual surplus over the sufficiency cut-off is not considered and do not contribute to a higher level of individual and general happiness: this facet is very interesting and recalls the works of Angus Deaton, recently awarded with the Nobel Prize\footnote{The Nobel Prize 2015 was awarded to Angus Deaton ``for his analysis of consumption, poverty, and welfare''.}, who identified, in a paper with Daniel Kahneman\footnote{He was awarded the 2002 Nobel Memorial Prize in Economic Sciences, shared with Vernon L. Smith.}, the income happiness threshold at \$\,75,000 \citep{kahneman_deaton_2010}.
This kind of social indicators have a number of problematic features. They have not solid theoretical foundations and in most cases they are presented without any framework for a rational discussion concerning results, meaning and construction (i.e. how the dimensions are correlated and - where aggregated- weighted). Another limitations refers to the scarce attention paid at the correlations among the various domains and proxies: aggregating them to create the individual well being index is to the detriment of the correlations between social dimensions at individual level \citep{fleurbaey_2009}.
Nevertheless, Stiglitz Commission Report had a strong impact in well being research giving institutional acknowledgement to the usefulness of self-reported measure of well being in the measurement of welfare as well as in public and social policy making. Indeed, in the last years, in spite of the traditional discretion of economists, the assessment of well being in a more direct way, from measuring circumstances to involving people, seems to be even more feasible and reliable thanks to surveys and questionnaires and a large amount of works exploiting them in various fields of research. Apart from surveys, other methods to evaluate individual wealth have been developing, such as experience sampling and day reconstruction method, physiological and neurological measures, behavioural observation \citep{fleurbaey_2009} and, more recently, social networks \citep{quercia_2015}.

\section{Perceived Well Being and Surveys Studies}
\label{sec.4}
In literature, subjective well being has been presented as a multifaceted concept. According to some scholars \citep{lyubomirsky_etal_2005} the notion of subjective well being includes dimensions of life satisfaction, usually defined in relation to different life domains, and levels of positive and negative affect: the first is defined as the evaluative or cognitive component of well being, while the second as the emotional component \citep{diener_1984, diener_etal_1999}. Other researchers \citep{ryan_deci_2001, deaton_stone_2013, steptoe_etal_2015} have defined subjective well being as threefold: there is an \textit{evaluative} domain of well being, that means life satisfaction, a \textit{hedonic} one, covering feeling of happiness, sadness, anger, illness and pain, and finally \textit{eudaimonic}, referring to the purpose and meaning in life. Despite all the different definitions, the methodologies put in place to assess the different components of subjective well being are surprisingly the same: research relies on the use of self-reported measures \citep{wojcik_2015}.
Among all the surveys used to study subjective well being, one can consider different kinds of surveys: surveys of a general nature that are made worldwide (Gallup World Poll, World Database of Happiness, World Values Survey), surveys of general nature that have local impact (Gallup-Healthways Well Being Index, British Household Panel Survey, European Social Values Survey(ESS), Eurobarometer, Global Health \& Wellbeing Survey), surveys that consider only certain groups of people, as youth (National Child Development Survey, Survey of Wellbeing of Young Children (SWYC)), students or employers (Gallup's surveys of workers and customers corporate clients,  Social-Emotional Wellbeing (SEW) Survey, GA Releases Graduate Student Happiness \& Well Being Report).
Angus Deaton use intensively both Gallup World Poll and Gallup Healthways Survey, the first sample is worldwide, while the second is US based. Among the several studies by Deaton on subjective well being, we provide here a couple of examples.

\subsection{Perceived Well Being and Income}
\cite{steptoe_etal_2015} distinguish evaluative well being, hedonic well being and eudaimonic well being, and present new analyses about the pattern of well being across ages and the association between well being and survival at older ages. Using the Gallup World Poll, a continuing survey in more than 160 countries, they showed a U-shaped relation between evaluative well being and age in high income, English speaking countries, with the lowest levels of well being in ages 45-54 years. However, they found that this pattern is not universal: respondents from Latin America also shows decreased well being with age, whereas well being in sub-Saharan Africa shows little change. They conclude that the well being of elderly people is an important objective for both economic and health policy.

\subsection{Well Being and Suicide Rates}
\cite{case_deaton_2015} relate well being measures to suicide rates. They used data from the United States and from other countries to examine patterns of suicide and well being by age and across space. They use data from the Gallup Healthways Wellbeing Index measures (2 million observations from 2008 through 2013) as well as data from the Gallup World Poll, which covers nearly all the countries in Europe, the OECD, and Latin America. They found differences in suicides between men and women, between Hispanics, blacks, and whites, between age groups for men, between countries or US states, between calendar years, and between days of the week, do not matched differences in life evaluation. Reports of physical pain were strongly predictive of suicide in many context. The prevalence of pain was increasing among middle-aged Americans, and was accompanied by an increase in suicides and deaths from drugs and alcohol poisoning.

\subsection{The Gallup World Poll}
The Gallup World Poll measures the attitudes and behaviours of the World's Residents. It tracks a lot of issues worldwide, such as food access, employment, leadership performance, and well being. Gallup senior scientists advise on the development of a common set of statistics Gallup collects in every country in the world. The survey includes more than 100 global questions as well as region-specific items. Gallup asks residents from Australia to Pakistan the same questions, every time, in the same way. This makes it possible to trend data from year to year and make direct country comparisons. Gallup uses telephone surveys in countries where telephone coverage represents at least 80\% of the population or is the customary survey methodology. Telephone methodology is typical in the U.S., Canada, Western Europe, Japan, Australia, etc. Gallup purchases telephone samples from various sample providers located in each region, including Sample Answers and Sample Solutions. In the developing world, including much of Latin America, the former Soviet Union countries, nearly all of Asia, the Middle East, and Africa, Gallup uses an area frame design for face-to-face interviewing in randomly selected households. Face-to-face interviews are approximately one hour, while telephone interviews are about 30 minutes. With some exceptions, all samples are probability based and nationally representative of the resident population aged 15 and older. The coverage area is the entire country including rural areas, and the sampling frame represents the entire civilian, non-institutionalized adult population of the country. Exceptions include areas where the safety of the interviewing staff is threatened and scarcely populated islands in some countries. The typical survey includes at least 1,000 individuals. In some countries, Gallup collects oversample in major cities or areas of special interest. Additionally, in some large countries, such as China and Russia, sample sizes of at least 2,000 are collected. Although rare, in some instances, the sample size is between 500 and 1,000. Gallup conducts the surveys on a semiannual, annual, and biennial frequency that is determined on a country-by-country basis.The questionnaire is translated into the major languages of each country. The translation process starts with an English, French, or Spanish version, depending on the region. A translator who is proficient in the original and target languages translates the survey into the target language. A second translator reviews the language version against the original version and recommends refinements. With the same survey they make three different well being index:

\begin{itemize}
\item The Thriving, Struggling, and Suffering Indexes measure respondents' perceptions of where they stand, now and in the future. Individuals who rate their current lives a ``7'' or higher AND their future an ``8'' or higher are ``Thriving.'' Individuals are ``Suffering'' if they report their current AND future lives as a ``4'' and lower. All other individuals are ``Struggling''.
\item The Positive Experience Index is a measure of experienced well being on the day before the survey. Questions provide a measure of respondents' positive experiences.
\item The Negative Experience Index is a measure of experienced well being on the day before the survey. Questions provide a measure of respondents' negative experiences.
\end{itemize}

\subsection{Well Being and Health: The Gallup-Healthways Wellbeing Index}
The Gallup-Healthways Wellbeing Index is the largest collection of data related to the health and wellbeing. The Gallup-Healthways Well Being Index began in January 2008, and surveys 1,000 Americans every day. The research and methodology underlying the Well Being Index is based on the World Health Organization definition of health as ``not only the absence of infirmity and disease, but also a state of physical, mental, and social wellbeing''. The Well Being Index measures Americans' perceptions of their lives and their daily experiences through five interrelated elements that make up well being: sense of purpose, social relationships, financial security, relationship to community, and physical health. Gallup interviews at least 500 U.S. adults aged 18 and older daily. More than 175,000 respondents are interviewed each year, and over 2 million interviews have been conducted to date since 2008. Since it began in 2008, the Gallup-Healthways Well Being Index survey has been conducted every day, excluding major holidays and other events, for 350 days per year. Gallup reports findings from this in weekly, monthly, quarterly, and yearly aggregates, and by region, state, and community, as appropriate on \url{Gallup.com}.
Gallup and Healthways in 2012 created the Gallup-Healthways Global Well Being Index to measure well being worldwide. Gallup added 10 questions to its World Poll in 2013, with each of the questions associated with one of the five elements of well being. The Global Well Being Index is an extension of more than six years of research and 2 million interviews in the U.S. through the Gallup-Healthways Well Being Index. The Global Well Being Index is a global barometer of individuals' perceptions of their well being and is the largest recent study of its kind. The Global Well Being Index is organized into the five elements:

\begin{itemize}
\item Purpose: liking what you do each day and being motivated to achieve your goals
\item Social: having supportive relationships and love in your life
\item Financial: managing your economic life to reduce stress and increase security
\item Community: liking where you live, feeling safe, and having pride in your community
\item Physical: having good health and enough energy to get things done daily 
\end{itemize}

In analyzing the results of the index, Gallup classifies responses as ``thriving'', well being that is strong and consistent, ``struggling'', well being that is moderate or inconsistent, or ``suffering'', well being that is low and inconsistent. The Global Well Being Index uses the same data collection and weighting methodology as the Gallup World Poll. Results for this Index are based on telephone and face-to-face interviews on the Gallup World Poll, with a random sample of approximately 133,000 adults, aged 15 and older, living in 135 countries and areas in 2013.

Each element in the Global Well Being Index contains two questions asked of all respondents:
\begin{itemize}
\item Purpose
\begin{itemize}
	\item You like what you do every day.
	\item You learn or do something interesting every day.
\end{itemize}
\item Social
\begin{itemize}
	\item Someone in your life always encourages you to be healthy.
	\item Your friends and family give you positive energy every day.
\end{itemize}
\item Financial
\begin{itemize}
	\item You have enough money to do everything you want to do.
	\item In the last seven days, you have worried about money.
\end{itemize}
\item Community
\begin{itemize}
	\item The city or area where you live is a perfect place for you.
	\item In the last 12 months, you have received recognition for helping to improve the city or area where you live.
\end{itemize}
\item Physical
\begin{itemize}
	\item In the last seven days, you have felt active and productive every day.
	\item Your physical health is near-perfect.
\end{itemize}
\end{itemize}
\cite{krueger_schkade_2008} analysed the persistence of various subjective well being questions over a two-week period for a sample of 229 working women. They found that both overall life satisfaction measures and affective experience measures derived from the DRM exhibited test-retest correlations in the range of .50-.70. It is surprising that measures intended to assess the general state of SWB over an extended period, such as overall life satisfaction, should not be more reliable than measures of affective experience on different days two weeks apart. One's general level of life satisfaction would be expected to change only very slowly over time, because most of the known correlates (age, income, marital status, employment) of life satisfaction also change slowly.

\subsection{Self-Rated Health Conditions and Well Being}
The \cite{sabatini_2014} analysis relies on a unique dataset collected through the administration of a questionnaire to a representative sample (n = 817) of the population of the Italian Province of Trento in March 2011. He tested the relationship between happiness and self-rated health in Italy. He found that happiness is strongly correlated with perceived good health, after controlling for a number of relevant socio-economic phenomena. Health inequalities based on income, work status and education are relatively contained with respect to the rest of Italy.

\subsection{The European Social Survey}
(\cite{fors_kulin_2015}) included affective well being in their analyses. Using European Social Survey data and multi-group confirmatory factor analysis, they estimate latent country means for the two dimensions, both life satisfaction and affective well being, and compare country rankings across them. The results reveal important differences in country rankings depending on whether one focuses on affective well being or life satisfaction. A limitation of their study is the use of retrospective assessments of affect.
The European Social Survey (ESS) is an academically driven cross-national survey that has been conducted every two years across Europe since 2001.
The survey measures the attitudes, beliefs and behaviour patterns of diverse populations in more than thirty nations. The main aims of the ESS are:
\begin{itemize}
\item to chart stability and change in social structure, conditions and attitudes in Europe and to interpret how Europe?s social, political and moral fabric is changing;
\item to achieve and spread higher standards of rigors in cross-national research in the social sciences, including for example, questionnaire design and pre-testing, sampling, data collection, reduction of bias and the reliability of questions;
\item to introduce soundly-based indicators of national progress, based on citizens? perceptions and judgements of key aspects of their societies;
\item to undertake and facilitate the training of European social researchers in comparative quantitative measurement and analysis;
\item to improve the visibility and outreach of data on social change among academics, policy makers and the wider public.
\end{itemize}
The ESS data are available for non-commercial use free of charge and can be downloaded from the website\footnote{\url{http://www.europeansocialsurvey.org/}} upon registration.
A key aim of the ESS has always been to implement high quality standards in its methodology and to improve standards in the field of cross-national surveys more generally. Measuring attitudes cross-nationally has challenges that go beyond those of surveys conducted in a single country or language.
In order to achieve ``optimal comparability`` of the ESS, the Core Scientific Team produces a detailed project specification, which is revised in light of each successive round. National teams of participating countries should read the specification in its entirety to ensure that fieldwork is conducted according to the same standards cross-nationally. This ``principle of equality or equivalence`` applies to sample selection, translation of the questionnaire, and to all methods and processes associated with data collection and processing.
The ESS has been collecting methodologically robust cross-national data on well being every two years since 2002. The survey includes headline measures of subjective well being such as ``life satisfaction`` and ``happiness`` as part of its core questionnaire, asked of respondents in each round. More in-depth data on well being is also provided for some rounds where thematic ``rotating modules``, which vary from 
round to round, have focused on different aspects of well being. These data on well being are collected alongside a large number of socio-demographic background variables and questions asking about other important social and political topics, providing a rich dataset.
In every ESS round, some questions have been included on subjective well being, social exclusion, religion, perceived discrimination and national and ethnic identity. In addition, some question measuring attitudes to immigration, that were originally fielded as part of the ESS1 (2002) rotating module on immigration, have been asked as part of the Core section from ESS2 (2004) onwards. 
The module that focus on the personal and social well being of respondents was first introduced in ESS3 (2006), and then repeated in ESS6 (2012) where the focus on both personal and social well being was retained. The ESS6 module also sought to incorporate a new validated scale of positive well being, and include questions to explore developments in the evidence base on well being promoting behaviours. Specific items focused on helping others, feelings in the last week, life satisfaction and physical activity.
The module that focus on the inter-relations between work, family and well being was first introduced in ESS2 (2004), and then repeated in ESS5 (2010) where it drew primarily on the ``work experience`` and ``work-family`` conflict sections of the previous double module, while retaining a number of key indicators with respect to household activity. Exploring these relations in a comparative perspective should add not only to a general understanding of sources of satisfaction and psychological strain among European populations, but also to the role of national welfare regimes in this process.
The module focusing on the social determinants of health and health inequalities was first fielded in 2014 (ESS7). Specific items included a range of health measurements (BMI, self-reported diagnoses and mental wellbeing); social determinants (childhood conditions, housing quality and working environment); behaviours (smoking, alcohol use, fruit and vegetable consumption and physical activity); and use of primary, secondary and alternative health care.
In particular it should be noted a new initiative\footnote{\url{http://esswellbeingmatters.org/}} ``Measuring and Reporting on Europeans Well Being``: Findings from the European Social Survey' which showcases the scope that ESS data provide for exploring the definition, distribution and drivers of subjective well being across Europe. Academics, policymakers and students are encouraged to explore the dedicated website and use the resource for their own research and informing policy.

\subsection{The RAND American Life Panel}
\cite{Kapteyn_etal_2015} used two waves of a population based survey (the RAND American Life Panel) to investigate the relations between various evaluative and experienced well being measures based on the English Longitudinal Study of Aging, the Gallup Well Being Index, and a 12-item hedonic well being module of the Health and Retirement Study. In a randomized set-up they administered several versions of the survey with different response scales. The RAND American Life Panel consisted of approximately 5,500 respondents ages 18 and over who were interviewed periodically over the Internet. They found that all evaluative measures load on the same factor, but the positive and negative experienced affect measures load on different factors. The relation of evaluative and experienced measures with demographics are very different; perhaps the most striking aspect is the lack of a consistent relation of experienced well being measures with income, while evaluative well being is strongly positively related with income. They found also evidence of an effect of response scales on both the estimated number of underlying factors and their relations with demographics.

\subsection{The World Values Survey}
\cite{sun_etal_2015} analysed how subjective well being in a Chinese population varies with subjective health status, age, sex, region and socio-economic characteristics. In the Household Health Survey 2010, face-to-face interviews were carried out in urban and rural counties in eastern, middle and western areas of China (n = 8,000, aged 15-102 years). To measure subjective well being, a validated Chinese version of a question on self-reported happiness, adopted from the World Values Survey, was used. They found that subjective well being increased with socio-economic status (income and education), and was lower among unemployed individuals and divorced individuals. But it also increased strongly with subjective health status. The reported subjective well being was also higher in rural counties than in urban counties in the same area.

The World Values Survey\footnote{\url{http://www.worldvaluessurvey.org}} (WVS) is a global network of social scientists studying changing values and their impact on social and political life, led by an international team of scholars. The survey, which started in 1981 limited to developed societies , nowadays consists of nationally representative surveys conducted in almost 100 countries which contain almost 90 percent of the world?s population, using a common questionnaire.
A second wave of WVS surveys was carried out in 1990/91 and a third wave in 1995/97, this time in 55 societies and with increased attention being given to analyzing the cultural conditions for democracy. A fourth wave of surveys was carried out in 1999/2001 in 65 societies. A key goal was to obtain better coverage of African and Islamic societies, which had been under-represented in previous surveys. A fifth wave was carried out in 2005?07 and a sixth wave was carried out during 2011/12.
The WVS is non-commercial, cross-national and time series investigation of human beliefs and values, currently including interviews with almost 400,000 respondents. Moreover the WVS  covers the full range of global variations, from very poor to very rich countries, in all of the world?s major cultural zones. The WVS measures, monitors and analyses: support for democracy, tolerance of foreigners and ethnic minorities, support for gender equality, the role of religion and changing levels of religiosity, the impact of globalization, attitudes toward the environment, work, family, politics, national identity, culture, diversity, insecurity, and subjective well being. 

\subsection{European Quality of Life Survey}
The purpose of the \cite{soukiazis_ramos_2015} study is to analyze the determinants of life satisfaction and happiness of the Portuguese citizens using data from the European Quality of Life Survey. They used micro-data to measure subjective well being by means of self-reported answers to a life satisfaction and happiness status. They found that trust in public institutions, satisfaction with material conditions, volunteering activities and employment status have a positive and significant effect on life satisfaction. Their evidence also shows that satisfaction with family, satisfaction with material conditions, participation in sport activities, optimism and the marital status are relevant factors in explaining citizen's happiness in Portugal.

\subsection{Potential Bias in Self-Reported Well Being}
As it is easy to see, even from the works just reported, self-reports are extensively used to study well being, forgetting they are a potentially misleading  source of information. Reports of well being are influenced by manipulations of current mood and of the immediate context, including earlier questions on a survey that cause particular domains of life to be temporarily salient \citep{schwarz_1999, schwarz_strack_1999}. Satisfaction with life and with particular domains is also affected by comparisons with other people and with past experiences \citep{clark_2003}. To overcome these biases \cite{kahneman_etal_2004} suggested that measures of well being must have the following characteristics: 
\begin{enumerate}
\item  they should represent actual hedonic and emotional experiences as directly as possible; 
\item they should assign appropriate weight to the duration of different segments of life; 
\item they should be minimally influenced by context and by standards of comparison.
\end{enumerate}
In the same work the authors proposed these following procedures. The Experience Sampling Method (ESM), where they collect information on individuals' experiences in real time in their natural environments, it is carried out with an electronic diary that beeps at random times during a day and asks respondents to describe what they were doing just before the prompt. Unfortunately it is not a practical method for national well being accounts, because above all it is impractical to implement in large sample; and infrequent activities are only rarely sampled. The Daily Reconstruction Method (DRM), they ask respondents to fill out a diary corresponding to events of their previous day. However, there still remains a problem: it  involves a retrospective report. The Event Recall Method (ERM), where they ask questions about feelings associated with particular events. In two of the same scholars proposed the $U$-index, a misery index of sorts, which measures the proportion of time that people spend in an unpleasant state, and has the virtue of not requiring a cardinal conception of individuals' feelings. Another difficulty of using data on subjective well being is that individuals may interpret and use the response categories differently. To solve this problem survey researchers try to anchor response categories to words that have a common and clear meaning across respondents, but there is no guarantee that respondents use the scales comparably. Despite the apparent signal in subjective well being data, one could legitimately question whether one should give a cardinal interpretation to the numeric values attached to individuals' responses about their life satisfaction or emotional states because of the potential for personal use of scales \citep{kahneman_krueger_2006}. Despite all the efforts made in the literature, and partly presented above, it remains much uncertainty in the use of self-reported data. Also a major user as Angus Deaton manifests these uncertainties \citep{deaton_2012}. After analzying Gallup data on 1000 Americans through the economic crisis of 2008 to 2010, he issued a strong warning on the use of subjective well being questions for cross-national comparisons. Although in the early days of the crisis the respondents reported lower levels of life satisfaction and greater anxiety, these measures had largely recovered by the end of 2010, despite the fact that high levels of unemployment indicated that the crisis was ongoing \citep{deaton_2012}. Although the measures picked up people's anxiety, they were no guide to the real state of the economy. Most devastatingly, however, the author reports that the order in which questions were asked (and specifically shifting questions about politics to just before the questions on life evaluation) `dwarfs' the effect of the crisis even at its worst. Although he endorses the use of subjective well being measures in showing up cross-sectional differences in life circumstances, \cite{deaton_2012} cautions that ``they still have a long way to go in establishing themselves as good time-series monitors for the aggregate economy''. He concludes the article: ``In a world of bread and circuses, measures like happiness that are sensitive to short term ephemera, and that are affected more by the arrival of St Valentine's Day than to a doubling of unemployment, are measures that pick up the circuses but miss the bread.'' 
Although the accumulating knowledge about the process underlying self-reports that in these years  promised to improve questionnaire design, a lot of issues are still open, and new alternatives or additions to surveys are improving .

\section{Social Networks Sites, Big Data and Well Being}
\label{sec.5.1}

It has it has been argued \citep{ditella_macculloch_2008} that economists typically measure what people do rather than listening to what people say. Nevertheless, the increasing of social interaction through digital devices and the remarkable progresses in statistics and computational science are changing the main habits of social scientists, providing them with both new tools and new kind of dataset with whom to study societies. 
Social Network Sites (SNS) host an enormous amount of records and digital interactions  that can be collected and analyzed for research purposes, making it possible to study social dynamics from an unseen perspective \citep{pentland_2014}. 
Thanks to the development in computational science and statistical theory, social sciences are fostering their capacity to manage and analyze set of data with unseen dimension and capillarity \citep{king_2011, lazer_etal_2009}.

\subsection{Why SNS? Asking vs Listening}
Sentiment or opinion analysis is the core aspect of a brand new method for measurement of happiness and well being. This research field is largely dedicated to the systematic extraction of web users' emotional state from the texts they post autonomously on different internet platforms, such as blogs, forums, social networks (e.g.,Twitter or Facebook) \citep{kramer_2010, ceron_etal_2013, curini_etal_2015}. 
Social psychologists have found a link between well being of individual and their use of words  it is possible to extract words from tweets allowing to reconstruct the emotional content the author wanted to communicate, to infer psychological traits and to measure well being \citep{quercia_2015}.

The availability of these large data sets have driven up the growth of theories and methodologies for  sentiment or opinion analysis. Despite many limitations \citep{Couper}, if correctly performed,  sentiment analysis seems to be a useful framework to exploit when the constraints of standard survey methodology may be too strong \citep{iacus2014, king_2015}. On one hand, in fact, there is no need for asking questions to the target population: all that the analyst has to do is to listen to the on-line conversations and classify the opinions expressed accordingly; on the other hand, the available information is updated in real time and hence the frequency of the well being evaluation can be as high as desired, theoretically allowing for separating the volatile and emotional component from the permanent and structural one.

\subsection{iHappy and Other Indexes of Happyness over Twitter}
There exists a wide set of works aiming at tracking happiness through Twitter. Generally, they employ automated sentiment analysis over a great number of tweets, in order to elaborate an indicator of happiness and then identify possible determinants and influential factors. 

Some researchers have proposed dictionaries or list of words useful to carry out sentiment analysis. \cite{dodds_etal_2011} studied happiness with real-time, remote-sensing, and unobtrusive text-based hedonometer applied over 46 billion words contained in nearly 4,6 billion expressions posted over a 33 month by 63 million unique users. \cite{schwartz_etal_2013} examine tweets from 1300 different US counties, measuring life satisfaction through the recurrence of words used and their topics, defined as sets of recurring words referring to a specific domain (for example outdoor activities, feelings, engagement). This language analysis is found to be predictive of the subjective well being of people living in those counties as measured by representative surveys and the topics identified provide a more detailed behavioural and conceptual insights on counties' life satisfaction than the usual socio-economic and demographic variables.

\cite{kramer_2010} calculates a ``Gross national happiness'' index, operationalized as the standardized difference between the use of positive and negative affect words, aggregated across days, through a sentiment analysis over approximately 100 million Facebook users since September of 2007. He then develops a graph, linked to the index score, that updates automatically every day, with a delay of two days. \cite{durahim_coskun_2015} compare ``official'' statistics of well being (measured with surveys on a province basis by the Turkish Statistical Institute) with the results of a social media analysis (led on 35 million tweets published in 2013 and in the first quarter of 2014) and find high similarity among them. \cite{abdullah_etal_2015} develop a Smile Index as indicator of happiness from Twitter, validated detecting smiles ``:)''  and in 9 million geo-located tweets over 16 months and compared with both text-based techniques and self-reported happiness. Their study explores temporal trends in sentiment, relates public mood to societal events and predicts economic indicators.

Events are seen to be one of the more impacting aspects on Twitter happiness. \cite{curini_etal_2015} propose the iHappy indicator, measured with an innovative statistical techniques on 43 millions of tweets posted daily during 2012 in the 110 Italian provinces. They find that the quality of institutions influences marginally the happiness level, while meteorological variables and extemporaneous events, including the variability of the spread between German and Italian Bonds, have the largest impact. The important role of events in influencing happiness and public mood have been noticed also by \cite{bollen_etal_2009}. They carry out a sentiment analysis of all tweets published on Twitter in the second half of 2008 through a psychometric instrument to extract six mood states (tension, depression, anger, vigour, fatigue, confusion). Then, they elaborate a multi-dimensional mood vector for each day in the time line finding that social, political, cultural and economic events do have a significant, immediate and highly specific effect on the six dimensions. Again concerning sentiment analysis and events detection \cite{sakaki_etal_2010} show as tweets analysis can effectively contribute to estimate the centers of earthquakes and accelerate focused intervention. After investigating the real-time interaction of catastrophic events in Twitter, they propose an algorithm to monitor tweets in order to detect this kind of events. They classify tweets on the basis of keywords, the number of words and their context and developed a probabilistic spatio-temporal model for identifying the event location.
Social scientists have also developed instruments to collect information from users' profile and their networks of followers and correlate them with official socio-economic indicators. \cite{bliss_etal_2011}, in a study based on nearly 40 million message pairs posted to Twitter between September 2008 and February 2009, employ a recently developed hedonometer to investigate patterns of sentiment expression through the social network structure. They find that users' average happiness scores is positively and significantly correlated with those of users one, two, and three links away and that more connected users write happier status updates. Also \cite{bergsma_etal_2013}, answering to a growing interest in automatically classifying social media users by various qualities, define an algorithm that extracts attributes (names and user-provided locations) from social media accounts of millions of Twitter users. Then they form clusters from which they accurately assign numerous often unspecified properties such as race, gender, language and ethnicity.
A great part of Twitter happiness research is focused on exploring the potentials of georeferencing technologies and their usefulness in measuring public mood and well being. \cite{cranshaw_etal_2012} introduce a clustering model and a research methodology useful to study the structure and the composition of a city through the analysis of 18 million location check-ins carried out by users of an online social network. They identify clusters as representations of independent dynamic areas of the city, validating them through qualitative analysis, and underline their relevance from the policy-maker perspective, fostering a higher degree of targetization in urban policies \citep{bollen_etal_2009,bliss_etal_2011}.

\subsection{Happyness and Deprivation}
\cite{quercia_etal_2012} are more focused on well being and happiness: they consider Twitter users based in many London census communities, and study the relationship between sentiment expressed in tweets and community socio-economic well being (i.e. the Index of Multiple Deprivation supplied by the UK Office for National Statistics). They find that the two are highly correlated: the higher the normalized sentiment score of a community's tweets, the higher the community's socio-economic well being. \cite{frank_etal_2013} analyse over 37 million geo-located tweets to characterize the movement patterns of 180,000 individuals employing a hedonometer: interestingly they find that expressed happiness increases logarithmically with distance from an individual's average location. Finally, \cite{mitchell_etal_2013} mix language analysis with tweets georeferencing: they examine 80 million words generated in 2011 on Twitter and collect socio-demographic characteristics of all US states and close to 400 urban populations. They elaborated classifications of states and cities based on their similarities in word usage and connected word choice and message length with urban features such as education levels and obesity rates. Then, they estimated the happiness levels of states and cities correlating it with demographic data: they find that happiness within the US was found to correlate strongly with wealth, showing large positive correlation with increasing household income and strong negative correlation with increasing poverty and obesity. They also find that a significant driver of the happiness score for individual cities was found to be frequency of swearing words.

\section{A New Social Well Being Index (SWBI)}
\label{sec.5}
As said, sentiment analysis is the core aspect of mood extraction from SNS. Here we propose to apply iSA\footnote{The  iSA technique \citep{ceron_etal_2015} has also been previously used to build the iHappy index \citep{curini_etal_2015}.} (integrated Sentiment Analysis)  method, explained briefly in Section \ref{sec10}, to derive a set of indicators of subjective well being that capture different aspects and dimensions of individual and collective life. The indicators are summarized in a global index named Social Well Being Index (SWBI) but each component can also be analyzed separately in relation with other well being measures. The term ``social" emphasizes that:
\begin{itemize}
\item the indicator monitors the subjective well being expressed by the society through the social networks;
\item SWBI is not the result of some aggregation of individual well being measurements: as it will be clear in what follows, the index directly measures the aggregate composition of the sentiment throughout the society.
\end{itemize}

\subsection{Aspects captured by the SWBI}
The eight indicators we evaluate concern three different well being areas: personal well being, social well being, well being at work. To be comparable with a composite well being index currently available through periodical surveys for the main European countries, we adopt the definitions introduced by the think-tank NEF (New Economic Foundation). Each well being area is analyzed by a single component and each component is defined through the hypothetic question one might find in a questionnaire \citep{nef_2012}. Let us point out once more that, in our case, these questions are just ``hypothetic": no explicit question can be submitted to the target population in our research, the sentiment and any kind of opinion are extracted from the text through the supervised analysis of the language used in the posts. The data source are tweets written in Italian language and from Itally and data are accessed through Twitter's public API. A small part of these data (around 1- to 5\% each day) contain geo-reference information which is used to build the SWBI indicator at province level in Italy. We have stored and analyzed more than 143 millions of tweets, about 100 thousands per day, of which only 1.2 millions of tweet are geo-localized at province level (about 1\%).

Here is the definition of each single components of SWBI:
\begin{enumerate}
\item {\it Personal well being}:
\begin{itemize}
	\item {\bf emotional well being}: the overall balance between the frequency of experiencing positive and negative emotions, with higher scores showing that positive emotions are felt more often than negative ones ($\tt emo$);
	\item {\bf satisfying life}: having positive evaluation of one's life overall ($\tt sat$);
	\item {\bf vitality}: having energy, feeling well-rested and healthy, and being physically active ($\tt vit$);
	\item {\bf resilience and self-esteem}: a measure of individual psychological resources, optimism and ability to deal with life difficulties ($\tt res$);
	\item {\bf positive functioning}: feeling free to choose and having the opportunity to do it; being able to make use of personal abilities and feeling absorbed and gratified in activities ($\tt fun$);
\end{itemize}
\item {\it Social well being}:
\begin{itemize}
	\item {\bf trust and belonging}: trusting other people, feeling to be treated fairly and respectfully and feeling sentiments of belonging ($\tt tru$);
	\item {\bf relationships}: extent and quality of interactions in close relationships with family, friends and others who provide support ($\tt rel$);
\end{itemize}
\item {\it Well being at work}:
\begin{itemize}
	\item {\bf quality of job}: feeling job satisfaction, satisfaction with work-life balance, evaluating the emotional experiences of  work and work conditions ($\tt wor$).
\end{itemize}
\end{enumerate}
As it is not possible to ask questions in social media, the components of the SWBI are obtained through the reading of a sample of tweets (see next Section for details) and trying to classify each tweet according to the scale -1, 0, 1, where -1 is for negative , 0 is neutral and 1 is positive feeling. For example, a text like ``I am grateful to my friends and relatives who sustained me during my hard times", will be classified as $\tt rel = +1$.
While a text like ``you can't really trust anyone nowadays", will be classified as $\tt tru = -1$; or a text like ``ok, let's go to work again today" as $\tt wor = 0$. These are of course just examples of how one derives the indicators from qualitative text analysis.

\subsection{The SWBI Behavior 2012-2015}
\label{sec:4.0}
The SWBI index is the simple arithmetic mean\footnote{We use simple mean here for sake of simplicity: any reasonable and justified weighted mean of the eight indicators can be theoretically proposed as a synthetic well being measure.} of the eight indicators $\tt emo$, $\tt fun$, $\tt rel$, $\tt res$, $\tt sat$, $\tt tru$,  $\tt vit$ and $\tt wor$ introduced in the above.
Table \ref{tab:all} reports the yearly values of SWBI and its eight components. Data are available from February 2012 till November 2015 at the time of this writing. The analysis is based on a total of 143 millions of tweets posted in Italian and from Italy. 

\begin{figure}[t!]
\centering
	\includegraphics[height=0.4\textheight]{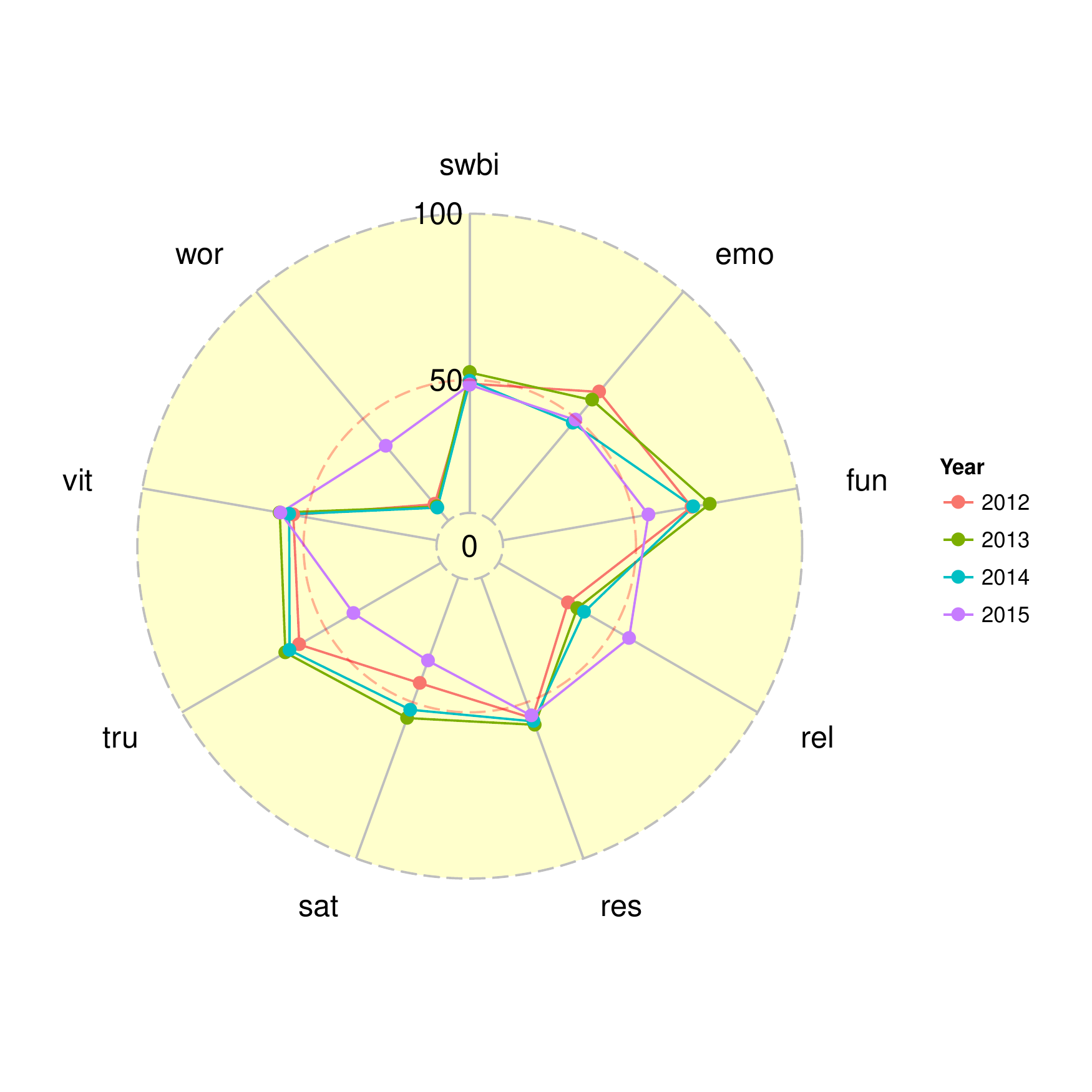}
\caption{\small Yearly average values of SWBI and its components. Data from Table \ref{tab:all}.}
\label{fig:radial}
\end{figure}

\begin{table}[ht]
\centering
\begin{tabular}{rrrrrrrrrr}
  \hline
 & SWBI & emo & fun & rel & res & sat & tru & vit & wor \\ 
  \hline
2012 & 48.87 & 60.55 & 67.76 & 34.10 & 55.10 & 43.88 & 59.22 & 53.91 & 16.44 \\ 
  2013 & 52.22 & 57.32 & 73.31 & 37.35 & 57.19 & 55.03 & 64.04 & 58.04 & 15.50 \\ 
  2014 & 49.69 & 48.24 & 68.26 & 39.73 & 56.11 & 52.37 & 62.59 & 55.15 & 15.10 \\ 
  2015  & 48.50 & 49.50 & 54.57 & 55.35 & 54.30 & 36.72 & 40.40 & 57.81 & 39.33 \\ 
   \hline
\end{tabular}
\caption{Average values of SWBI and its components.}
\label{tab:all}
\end{table}
It is interesting to notice that, if we look at the per capita GDP in Italy in 2012-2014 (data for 2015 are not available yet) and the value of the corresponding SWBI indicator we cannot find a clear common path, meaning that there is not necessarily a direct relationship between the level of economic activity of the country and the perceived well being. The well being indicator, in other terms, does not seem to simply reflect the conditions of the economic system, even in a period of serious economic crisis.  \begin{center}
\begin{tabular}{c|ccc}
Year & 2012 & 2013 & 2014\\
\hline
SWBI & 48.87& 52.22 & 49.69\\
\hline
GDP per capita (in euros, curr. prices) & 26760.0 & 26496.1 & 26545.8\\
\end{tabular}
\end{center}
\begin{figure}[t]
\centering
	\includegraphics[width=\textwidth,height=0.3\textheight]{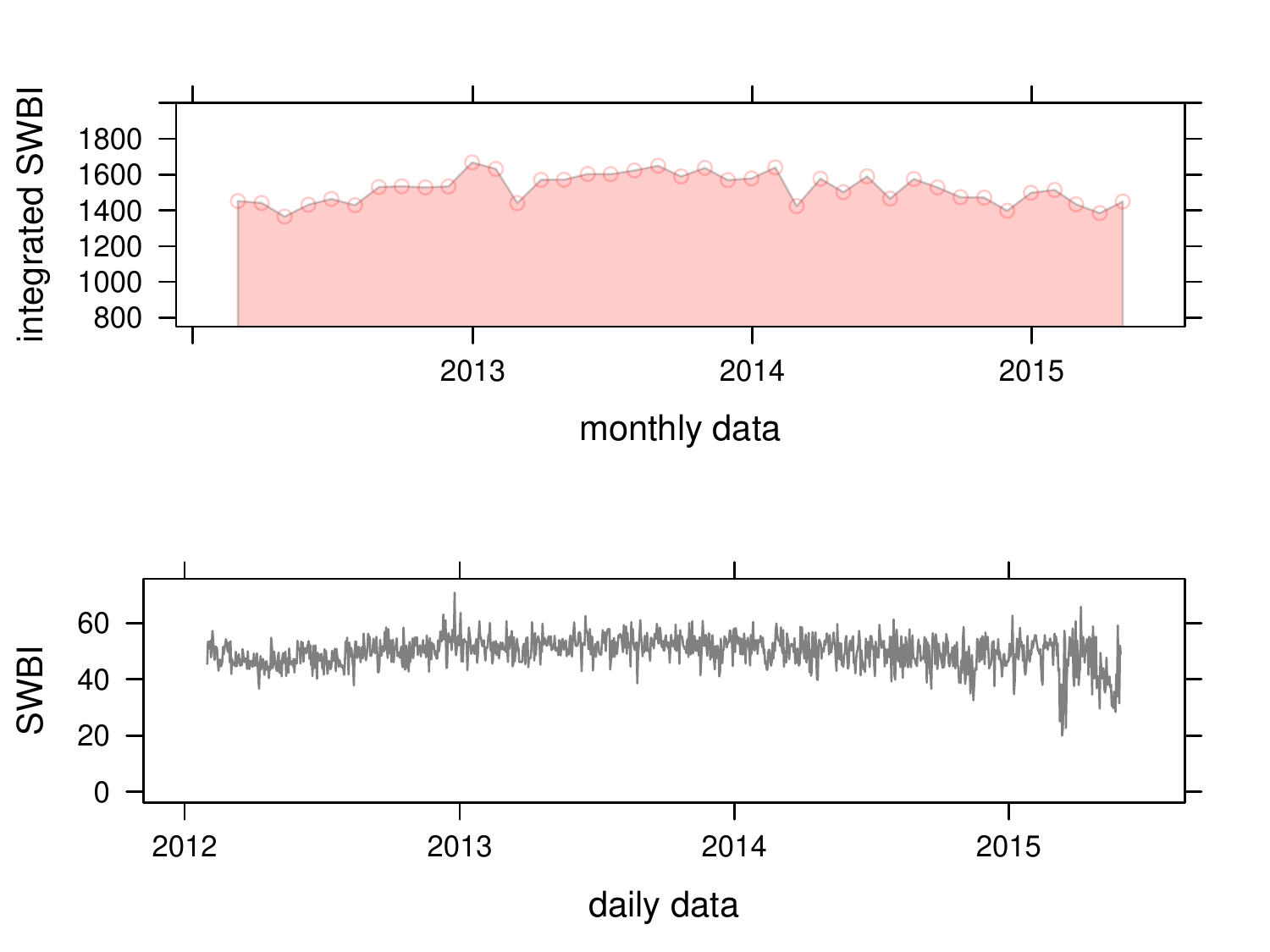}
\caption{\small daily values of SWBI (bottom panel) and its monthly integrated value (upper panel). The integrated value of SWBI represents the gross balance of well being during each period.}
\label{fig:swbi}
\end{figure}
Figure \ref{fig:radial} represents the same data as Table \ref{tab:all}. It is easy to note that the values of the indicator in 2015 show remarkable differences (both positive and negative) compared to the trend of the previous years: see, in  particular, the increase in $\tt wor$ and $\tt rel$, or $\tt tru$, $\tt fun$ and $\tt sat$ for the opposite variation.
In addition to that, Figure \ref{fig:swbi} contains the plot of the daily values of SWBI (bottom panel) and its monthly integrated value (upper panel). The integrated value of SWBI represents the gross balance of well being during each period. This representation dumps down the irregularity and high variability of daily estimates, which is typical in social media data. The above descriptive statistics need an in depth evaluation. It is only the case to note that the indicator registers both structural and volatile components of well being and what we are showing is a preliminary and rough separation of the two, which is one of the discussion topic in the literature on well being measurement. 

\section{Auxiliary results: iSA technique}
\label{sec10}
We briefly present the iSA algorithm which is used in the construction of the SWBI components of Section \ref{sec.5}.
iSA is a human supervised statistical method, where part of the texts are read by humans and part is classified by the machine. The supervised part is essential in that this is the step where information can be retrieved from texts without relying on dictionaries of special semantic rules. Human just read a text and associate a topic or opinion (for example: $D$ = ``satisfied at work'') to it. Then, the computer learn the association between the whole set of words used in a text to express that particular opinion and extends the same rule to the rest of the texts to be analyzed.

More formally, let us denote by $\mathcal D=\{D_0, D_2, \ldots, D_M\}$ the set of possibile categories (i.e. sentiments or opinions). The target of interest is $\{P(D), D\in\mathcal D\}$, i.e. the distribution of opinions in a corpus of $N$ texts. Normally, $D_0$ refers to  the texts corresponding to Off-topic or not relevant texts with respect to the analysis (i.e. the \textit{noise} in this framework).
Let $S_i$, $i=1, \ldots, K$, be a unique vector of $L$ possible stems (i.e. single words, unigrams, bigrams, etc which remain after the stemming phase) which identifies one of the texts in a corpus. More than one text in the corpus can be represented by the same $S_i$ and is such that each element of it is either $1$ if that stem is contained in a text, or $0$ in case of absence. The data set is then formalized as the set $\{ (s_j, d_j), j=1, \dots, N\}$ where $s_j\in\bar{\mathcal  S}$ (the space of possible vectors $S_j$) and $d_j$ can either be ``NA'' (not available or missing) or one of the hand coded categories $D\in \mathcal D$.

The ``traditional'' approach  includes all machine learning methods and statistical models that: 
\begin{enumerate}
\item use the individual hand coding from the training set to construct a model $P(D|S)$ for $P(D)$ as a function of $S$,  e.g. multinomial regression, Random Forests (RF), Support Vector Machines (SVM) etc.; 
\item predict the outcome of $\hat d_j=D$ for the texts with $S=s_j$ belonging to the test set; 
\item  when all data have been imputed in this way, these estimated categories $\hat d_j$ are aggregated to obtain a final estimate of $\hat P(D)$.
\end{enumerate}
 In matrix form, we can write
\begin{equation}
\underset{M\times 1}{P(D)} =  \underset{M\times  K}{P(D|S)} \underset{ K\times 1}{P(S)}
\label{eq1}
\end{equation}
where $P(D)$ is a $M\times 1$ vector, $P(D|S)$ is a $M\times  K$ matrix of conditional probabilities and $P(S)$ is a $ K\times 1$ vector which represents the distribution of $S_i$ over the corpus of texts. Once $P(D|S)$ is estimated from the training set with, say, $\hat P(D|S)$, then for each document in the test set with stem vector $s_j$, the opinion $\hat d_j$ is estimated with the simple Bayes estimator as the maximizer of the conditional probability, i.e. $\hat d_j= \arg\max_{D\in\mathcal D} \hat P(D|S=s_j)$. As it is well known, the present approach does not work if $P(D_0)$ is very large compared to the rest of the $D_i$'s.
iSA \citep{ceron_etal_2015} follow the idea by \cite{hopking} of changing the point of view but goes one step further in terms of computational efficiency and variance reduction.
Instead of equation \eqref{eq1}, one can  consider this new equation
\begin{equation}
\underset{  K\times 1}{P(S)} =  \underset{  K\times M}{P(S|D)} \underset{M\times 1}{P(D)}
\label{eq2}
\end{equation}
where now $P(S|D)$ is a $  K\times M$ matrix of conditional probabilities whose elements $P(S=S_k|D=D_i)$ represent the frequency of a particular stem $S_k$ given the set of texts which actually express the opinion $D=D_i$. 
Then, the solution of the problem is as follows
\begin{equation}
\text{(inverse problem)}\hspace{0.5cm}
\underset{M\times 1}{P(D)} = \underset{M\times M}{[P(S|D)^T P(S|D)]}^{-1} \underset{M\times   K}{P(S|D)}^T \underset{  K\times 1}{P(S)}
\label{eq3}
\end{equation}
Equation \eqref{eq3} is such that the direct estimation of the distribution of opinion $P(D)$ is obtained but individual classification is no longer possible. In fact, this is not a limitation as the accuracy of \eqref{eq3} with respect to \eqref{eq1} is vastly better (variance of estimates decreases from 15-20\% to 3-5\%). Moreover, researchers are comprehensibly more interested in the aggregate distribution of opinions throughout a population than in the estimation of individual opinion. For complete details see \cite{ceron_etal_2015}.

\section{Conclusion}
\label{concl}
The evolution of methods for well being measurement has pointed out two main issues: a) measurement of well being through objective quantities (mainly, GDP) allows only for a partial survey of the actual conditions of a community and does not take in adequate account several aspects of individual and collective life that have an impact on the perceived level of welfare. This has shifted the research interest from objective indicators of welfare to subjective and self-reported measures of well being, giving rise to indicators of well being based on questionnaires and surveys where people are asked to declare their own perception of current well being or evaluation of life satisfaction; b) measurement of well being through questionnaires produces biased results, because putting explicit questions on happiness and well being has a significant impact on self-evaluations: people give answers that are conditioned by the content and the order of questions or even by the fact that someone is putting a question.

In other words, the literature on well being measurement seems to suggest that ``asking'' for a self-evaluation is the only way to estimate a complete and reliable measure of well being, but at the same time ``not asking'' is the only way to avoid biased evaluations. Here we propose a method for estimating the welfare perception of a community simply ``listening'' to the virtual conversations people hold on the web. Exploiting the big data sources of information made available by social networks, we have obtained a set of well being measures - summarized by the Social Well Being Index (SWBI) - that span the wide range of welfare dimensions. The extraction of information from the big data repositories is allowed by a recent sentiment analysis technique, called iSA. The limited cost of acquisition of these information (compared to more traditional surveys), the continuous updating of data sources and the data process speed represent some of the advantages of these approach: for these reasons, the indicators are also good candidates for nowcasting analyses, i.e. for reducing the time leg between when an event occurs and when its consequences are known. The proposed methodology can be applied in other countries and different languages thanks to the adoption of the iSA algorithm.

\newpage

\bibliographystyle{spbasic}      
\bibliography{review}   

\end{document}